\def\kms{\ifmmode{\,\hbox{km}\,s^{-1}}\else {\rm\,km\,s$^{-1}$}\fi}

\def\hmpc{\ifmmode{h^{-1}\,\hbox{Mpc}}\else{$h^{-1}$\thinspace Mpc}\fi}

\def\eg{{\it e.g.}~}

\def\et{{\it et~al.}~}

\documentstyle[11pt,aaspp4]{article}
\tighten

\begin{document}

\title
{A Differential Volume-Redshift Test}

\author
{
R.~G.~Carlberg
}

\affil{Department of Astronomy, University of Toronto}
\authoraddr{Toronto, M5S 3H8, Canada, name@astro.utoronto.ca}


\begin{abstract} 
The geometry of Freedman-Roberston-Walker cosmological models is fixed
by the mass density parameter, $\Omega_M$, and the cosmological
constant, $\Omega_\Lambda$.  The classical volume-redshift
cosmological relation is a sensitive
${\bf\Omega}=[\Omega_M,\Omega_\Lambda]$ indicator but its redshift
dependence is observationally degenerate with the luminosity or
number density evolution of galaxies.  Introducing a measurement of
the invariant co-moving mass density of the universe reduces the
problem of galaxy evolution to a differential measurement between
clustered and field galaxies.  The cost is a 25\% reduction in
sensitivity to the $\bf\Omega$'s, although this test still remains
50\% more $\bf \Omega$ sensitive than the magnitude-redshift relation.
An implementation of the test as the product of the mass-to-light
ratio, $M/L$, of some clustered systems such as galaxy groups or
clusters, with $j/\rho_c$, the normalized luminosity density, is
considered. Over the zero to one redshift range the apparent
$\Omega_e(z)=M/L\times j/\rho_c$ has a zero point and slope related to
$\Omega_M$ and $\Omega_\Lambda$, respectively.  All quantities are
used in a differential sense, so that common selection effects,
dynamical scale errors, and galaxy evolution effects will largely
cancel.  The residual differential galaxy evolution between field and
the clustered galaxies can be measured from the sample data.  Monte
Carlo simulations, calibrated with observational data, show that 20
clusters spread over the zero to one redshift range, each having 100
cluster velocities, allows a 99\% confidence discrimination between
open and closed low density universe models. A similarly distributed
sample of about 100 rich clusters, or about 1000 galaxy groups found
within a large field survey, will measure $\Omega_\Lambda$ to about
7\% statistical error.
\end{abstract}

\clearpage
\section{Introduction}

The existence of a nonzero cosmological constant, $\Lambda$, would
have profound significance for our understanding of the universe and
its physics (\cite{weinberg}).  The expansion history and geometry of
the universe, as described by the Freedman-Roberston-Walker (FRW)
solution, are completely determined given the density parameter,
$\Omega_M=\rho_0 8\pi G/3H_0^2$, and the cosmological parameter,
$\Omega_\Lambda=\Lambda c^2/3H_0^2$. Knowledge of ${\bf
\Omega}=[\Omega_M,\Omega_\Lambda]$ is also of practical concern to
interpret the physical properties of objects at large redshifts. The
geometrical effects of the cosmological parameters are the basis of a
number of classical tests of the world model. These include the
redshift dependence of galaxy numbers, sizes or luminosities (\eg\
\cite{s61a,ppc}).  The success of any of these tests is in
large part dependent on the degree to which the evolution of the
intrinsic properties of galaxies is understood so that those effects
can removed to leave the cosmological variation of interest (\eg\
\cite{s61b,tinsley,s88,oh,tr}). The number count magnitude relation 
has long been taken as a hint that the $\Omega_M=1$ model might not be
correct, but this remains locked in the controversies of galaxy
formation (\eg\ \cite{ls,koo_kron,ldss,cowie,fukugita,sf}). Although both
the observational and theoretical understanding of galaxy evolution is
advancing rapidly the fundamental degeneracy between galaxy evolution
measurements and the cosmological parameters means that to derive a
reliable empirical model requires additional information.

There are a number of alternate approaches designed to establish
observational constraints on the value of $\Omega_\Lambda$.  One
geometrical test is to compare the redshift and angular extensions of
some physically understood shape at a large redshift, such as the
quasar-quasar correlation function (\cite{ap,qsos1,qsos2,qsos3}).  The
optical depth for multiple gravitational images of distant quasars
increases rapidly with positive values of $\Omega_\Lambda$. The
relatively low frequency of split images argues that
$\Omega_\Lambda\lesssim 0.7$ (\cite{cpt,mr,kochanek}). The high
precision of photometry coupled with the growing understanding of
supernovae, particularly those of type Ia (\cite{hamuy,riess}), allows
the classical magnitude-redshift test to be implemented.  With
sufficient data over a large redshift range
(\cite{perlmutter,schmidt}) both the $\bf \Omega$ values can be
measured.  Of particular note is the ``first Doppler peak'' in the
angular fluctuation spectrum of the Cosmic Background Radiation which
is a measurement over the largest possible path length of the geometry
of the universe (\eg\ \cite{bond} and references therein).

The cosmological parameters are of sufficient importance that they
will be measured with a variety of independent methods to establish
their values with confidence, understand the astrophysics of the
objects, and, to some degree test the FRW model itself.  The purpose
of this paper is to note a variant of the classical volume-redshift
test which breaks the cosmology-galaxy evolution degeneracy of the
volume-redshift relation.  That is, the co-moving mass density, which
is an invariant at low redshift in all conventional cosmologies, is
equal to the product of the total mass-to-light ratio, $M/L$, with the
field luminosity density, $j$ (\cite{oort}).  The virialized systems
can range from rich clusters to small groups of galaxies.  Virialized
systems have the considerable benefit that their mass profiles can be
inferred from dynamical techniques, independent of the distribution of
the galaxies.  Furthermore dynamical mass measurements are distance
independent (other than the cosmological factors of interest) with no
corrections required to compare masses at different redshifts.

The following section briefly reviews the volume-redshift relation in
the FRW models. The relations are expanded to first order to
illustrate the parameter dependencies and their degeneracy in the
luminosity function. The mass-density constraint is introduced and the
redshift dependence of the apparent $\Omega_e(z)$, which is a function
of the true $\bf \Omega$, is shown in section 3.  In Section 4 the
random errors and data requirements of practical measurements are
evaluated, concluding that a high precision measurement is primarily a
matter of assembling the appropriate datasets, which is likely to be
done anyway for a variety of other purposes.

\section{Distances and Volumes}

The co-moving distance, $r(z)$, to an object at redshift $z$ in an FRW
metric is found by integrating radially outward along the null
geodesic, $c \, dt = a(t)\,dr/\sqrt{1+r^2/R^2}$, where $a(t)=(1+z)^{-1}$
is the expansion factor and $R^{-2}= \Omega_R H_0^2/c^2$ (which is
positive in this metric for an open, negatively curved, universe). The
integral is rewritten in terms of the observable redshift using the
cosmological equation $H(z)=H_0 E(z)$, where $E^2(z)=\Omega_M(1+z)^3
+\Omega_R(1+z)^2+\Omega_\Lambda$ with
$\Omega_M+\Omega_\Lambda+\Omega_R\equiv 1$ (following the notation of
Peebles 1993). The resulting co-moving distance is,
\begin{equation}
r(z) = {c\over H_0} {1\over |\Omega_R|^{1/2}}
\hbox{\rm sinn}\,\Big[|\Omega_R|^{1/2}
	\int_0^z{ dz^\prime\over{E(z^\prime)}}\Big].
\label{eq:rco}
\end{equation}
The function sinn$(x)$ is $x$ for $\Omega_R=0$, $\sin(x)$ for
$\Omega_R<0$, and $\sinh(x)$ for $\Omega_R>0$ (\cite{cpt}).  The
co-moving volume element per $dz$ and per steradian is $r^2 \, dr/dz$,
which with the local Hubble law, $H(z)\,dr\, (1+z)^{-1} =
c\,dz\,(1+z)^{-1}$, becomes,
\begin{equation}
{{dV}\over{dz}} = {c\over H_0} {r^2(z)\over{E(z)}}.
\label{eq:vol}
\end{equation}

It is useful to note the first order expansions (relative to a
Euclidean background) of the co-moving distance,
\begin{equation}
r(z) \simeq {c\over H_0}z [1 + {1\over4}(-2-\Omega_M+2\Omega_\Lambda)z]
\label{eq:arco}
\end{equation}
and the co-moving volume element,
\begin{equation}
{{dV}\over{dz}} \simeq {c^3\over H_0^3} z^2[1+(-2-\Omega_M+2\Omega_\Lambda)z].
\label{eq:avol}
\end{equation}
The low redshift deceleration parameter is
$q_0=\Omega_M/2-\Omega_\Lambda$ and can be used to simplify these
first order expansions.  There are several noteworthy points to take
from these expressions. First, the volume-redshift relation has a
sensitivity to the $\bf \Omega$ parameters that is twice of the
magnitude-redshift, $(1+z)^2 r^2(z)$, relation. Second, the dependence
on the cosmological parameters of the distance and the volume element
is identical, meaning that luminosity evolution and geometry (or
density evolution) are degenerate at this order (which remains
approximately true over a wide range of redshifts). Furthermore the
test requires a comparison of objects at different redshifts. This
generally requires an absolute comparison of fully calibrated
quantities, along with all their selection effects.

\section{The Apparent $\Omega$}

Here we propose a test which is completely differential in the
observational quantities: galaxies are only compared to one another at
the same redshift and only redshift independent quantities (except for
the cosmological parameters of interest) are compared at different
redshifts.  The co-moving mass density of the universe is an invariant
for conventional cosmologies. One estimator of the mass density is
through Oort's method, $\Omega_e(z) = {M/ L}\times j/\rho_c$, where
$M/L$ is the total mass-to-light ratio of the universe and $j(z)$ is
the average field luminosity per unit volume.
In the interval $[z,z+\Delta z]$ 
and solid angle $\Delta \omega$, 
\begin{equation}
j(z)=4\pi (1+z)^2 r^2(z) {{\sum_{\Delta z\Delta \omega} f}
\over{\Delta z\Delta\omega \,dV/dz}},
\label{eq:j}
\end{equation}
where the $f$ are the observed fluxes of the field galaxies
in this volume.  The virial mass, or any other dynamical estimator of
the gravitational mass (including the mean lensing gravitational shear
within an aperture, \cite{ks}), is of the form
$M=3G^{-1}\sigma_1^2\theta_v r(z)(1+z)^{-1}$.  The quantity $\theta_v$
is the angular scale radius of the cluster, such as either the
classical pointwise virial radius estimator or a ringwise estimator
(\cite{pc,profile}).  The total cluster luminosity is $L= 4\pi (1+z)^2
r^2(z) \sum_c f $, where the sum adds the fluxes of the galaxies in
the redshift and angular range that define the cluster, and is limited
at the same $f$ or absolute luminosity as the field galaxies. The
apparent density parameter is then,
\begin{equation}
\Omega_e(z)= 
	\Big[ 
	{{\sum_{\Delta z\Delta\omega} f}\over{\Delta z\,\sum_c f}}
	{{3\sigma_1^2\theta_v}\over{G(1+z)\Delta\omega}}
	\Big]
	{H_0\over {c\rho_c}}{E(z,{\bf \Omega^i})\over 
	r(z,{\bf \Omega^i})},
\end{equation}
where ${\bf \Omega^i}$ are some convenient, but not necessarily
correct, values used to calculate the relation.  Note that all the
observational information is contained between the square brackets.
If we assume $\Omega_M=\Omega_M^i$, $\Omega_\Lambda^i=0$, the
resulting effective $\Omega_e(z)$ is then, to first order,
\begin{equation}
\Omega_e(z) \simeq \Omega_M [1 + {3\over 4}(\Omega_M^i
	-\Omega_M+2\Omega_\Lambda)z].
\label{eq:ao}
\end{equation}
The function $\Omega_e(z)$ has a zero point which gives $\Omega_M$ and
a redshift dependence ${3\over 2}\Omega_\Lambda z$, assuming that the
initial value, $\Omega_M^i$, is close to the true $\Omega_M$.  This
redshift dependence is 25\% less sensitive to $\Omega_\Lambda$ than
the volume-redshift test.

The general behavior of $\Omega_e(z)$ is shown in
Figure~\ref{fig:lamm}.  The plotted lines assume that the we
calculated the masses and luminosities using $\Omega_M^i=0.2$ and
$\Omega_\Lambda^i=0$.  For this choice of $\bf
\Omega^i$  the $\Omega_e(z)$ are functions of the true $\bf \Omega$ as,
\begin{equation}
\Omega_e(z) = 0.2{r(0.2,0,z)\over
		{r(\Omega_M,\Omega_\Lambda)}}
		{E(\Omega_M,\Omega_\Lambda)\over{ E(0.2,0,z)}}
\label{eq:lcurve}
\end{equation}
It is clear from this plot that the expected variation of this
quantity between redshift zero and unity if sufficient to extract both
$\Omega_M$ and $\Omega_\Lambda$. Moreover, the flat,
$\Omega_M+\Omega_\Lambda=1$, models have a distinctly different
behavior than open models, for low values of $\Omega_M$.

\section{Error Analysis}

A practical implementation depends on having sufficient data that the
random errors in the result are reduced to the desired level. The data
must also allow for checks for systematic errors, notably differential
evolution between clustered and field galaxies and whether clustered
systems have any segregation between their luminosity and their mass
distributions. 

\subsection{Random errors in $\Omega_e(z)$}

In the following analysis we will consider data which uniformly cover
the zero to unity redshift range, which is nearly optimal for the
application of this test. A smaller redshift range does not give much
leverage for the redshift dependence of $\Omega_e(z)$ which is
essential for $\Omega_\Lambda$ measurement.  On the other hand,
pushing the redshift range beyond redshift unity is quite difficult,
since many of the spectral features used to measure accurate
velocities, in particular the H+K lines and the 4000\AA\ break move
out of the region accessible to high efficiency optical spectrographs.

The random errors in estimating the $M/L$ ratio of a single virialized
cluster are straightforward to evaluate. For the dynamical estimator
of the form given above we need to estimate $\sigma_1^2$, $\theta_v$
and $L$. If the errors are uncorrelated then the fractional error of a
single cluster will be approximately $\sqrt{6/N}$. This expectation is
borne out quite accurately in available data (\cite{global}) in which
$N$ varies from about 25 to nearly 200. Once $N$ becomes much larger
than 100 the statistical error continues to decline in the expected
manner but the total error is dominated by projection effects and the
internal substructure of the cluster. Furthermore at about a magnitude
below $M_\ast$ in the cluster the field galaxies begin to overwhelm
the cluster galaxies. When these limits are encountered, it is better
to spread the observations over more clusters rather than continuing
to observe the same cluster. This is doubly true since more velocities
usually require observing more deeply into the luminosity function
where the fraction of the galaxies observed that are in the cluster,
as opposed to the field, is an ever declining fraction.  In summary,
it is readily feasible to obtain $M/L$ values of individual clusters
accurate to 25\%, which formally requires 96 cluster members
distributed over the face of the cluster. This is only practical for
very rich clusters of galaxies.

The number of clusters required for a confident measurement of
$\Omega_\Lambda$ is easily evaluated with Monte Carlo simulations of
the sample properties. Figure~\ref{fig:chilf} shows the results of
1000 simulations of a sample of 20 clusters that are randomly but
uniformly distributed over the $0\le z \le 1$ interval. The cosmology
used to generate the distribution has ${\bf \Omega}=[0.2,0.8]$ whereas
the $\Omega_e(z)$ are calculated assuming ${\bf \Omega^i}=[0.2,0]$. The
$1\sigma$ confidence range is $0.60\le \Omega_\Lambda \le 0.93$,
irrespective of $\Omega_M$. The 99\% confidence interval is $0.19 \le
\Omega_\Lambda
\le 1.09$, and many of the extreme values result primarily from unusually
poor random distributions in redshift, which could be readily avoided
in real observations.

To increase the precision of the result requires a larger cluster
sample.  One hundred clusters, with the same redshift and error
distribution as above, can reduce the error in $\Omega_\Lambda$ to a
7\% $1\sigma$ error.  Groups found in a field survey can also be used,
however, in that case the errors in $M/L$ for a single group are quite
large. An efficient use of the data will be to average the groups
together to build up pseudo-clusters of about 500-1000 galaxies which
decreases the substructure and projection effects to a level where the
profiles of galaxy density and mass can be checked for variation with
redshift as well as the changes in galaxy population with redshift.
Since between 1/10 and 1/3 of galaxies are in field groups this
implies that a field survey of about 20,000 galaxies is in hand.  In
either case, these surveys are easily feasible with existing
instrumentation and will become easier with new facilities.

\subsection{Differential Luminosity Evolution}

If there was no differential evolution between the galaxies used to
estimate $M/L$ and the average over the universe, then luminosity and
density evolution would have no impact on our measurement of
$\Omega_e(z)$.  The virialized systems that will be used range from
groups, which contain galaxies quite similar to field galaxies, to
rich clusters, which feature far more E and S0 galaxies, having redder
colors, than the field galaxy population.  Over the $z<1$ redshift
range under discussion all low redshift galaxies (of high surface
brightness) have a parent (or possibly several) at higher redshift
hence the accounting the accounting for mass and luminosity evolution
is not confused by completely new galaxies appearing in abundance.
For both luminosity evolution and density evolution the most difficult
parameter to determine is the characteristic luminosity,
$M_\ast$. Color differences, which track luminosity evolution
(\cite{lt}), can be measured to much higher precision.

The pioneering work on faint galaxy evolution established that the
brightest galaxies do not evolve much more than a minimal passive
evolution of (\cite{koo_kron,ldss}) as expected for bright cluster
galaxies (\cite{ble,smail,sed}).  Various observations continue to
bear out the basic slow evolution situation (\cite{schade_e,schade_d})
in spite of the significant changes of both cluster and field
populations at fainter absolute magnitudes
(\cite{bo,cfrs,ldss,lin,dressler,smail}).  The complications of
differential evolution can be greatly minimized by restricting the
sample to galaxies more luminous than about $M_\ast+1$
mag. Furthermore, by fully sampling the volume of the virialized
cluster or group one obtains a sample of cluster galaxies that is
closer to the field population, both in range of colors and
morphologies, than the central E/S0 galaxy population.

For measurements involving the comparison of galaxies in rich clusters
to the field there will be some differential evolution, which we
parameterize as $\Delta(j/L,z) =
\Delta^{jL}_0+z \Delta^{jL}_1$. As emphasized above, 
the $\Delta^{jL}$ are small compared to the evolution in the $j$ and
$L$ themselves, for suitably chosen samples.  The straightforward way
to determine the $\Delta^{jL}$ is through fitting the luminosity
functions, $M_\ast(z)=M_\ast(0)+z\Delta_M$ to cluster and field
galaxies individually, then $\Delta^{jL}_1$ is the difference between
$\Delta_M$ in clusters and the field.  It is important to note that
the values of the $\Delta_M$ will depend on the assumed $\bf \Omega$,
however the quantities are only used to find the difference in
characteristic magnitudes at the same redshift.  The $\Delta^{jL}$
have no $\bf \Omega$ dependence, being just flux ratios at a common
redshift.

The data gathered for the $\Omega_e(z)$ analysis will also be used to
measure the luminosity function relative to the field
galaxies. Although one should strive to make this an absolute
luminosity function with well defined sample criteria, its primary use
is in comparing clustered galaxies to field galaxies so sample
selection effects that are in common will make no difference to the
difference between the two luminosity functions.  The maximum
likelihood technique of Sandage, Tammann \& Yahil (1972).  A dataset
of 2000 absolute magnitudes is generated from a Schechter luminosity
function, $\phi(L,z)=(L/L_\ast)^{-\alpha}\exp{(-L/L_\ast)} $. As a
reasonable match to the available data $M_\ast(z) = M_\ast(0) +z
\Delta M_z $ mag, with $M_\ast(0)=-20$, $\Delta M_z=-0.7$ and
$\alpha=-1$. We assume that the data extend to $M=-19$ mag, although
the precise depth makes little difference providing it is 1 to 2
magnitudes below $M_\ast$. The 68, 90 and 99\% error ellipses are
shown for 3 parameter fits for $\alpha$, $M_\ast(0)$, and $\Delta M_z$
are shown in Figure~\ref{fig:chilf}. From this we conclude that the
sample will allow a measurement of $M_\ast$ accurate to 0.1 mag and
$\Delta M_z$ accurate to 0.2 mag per unit redshift. Normally the field
sample will be larger than the cluster sample, so the errors in
measuring the same quantities in the field will be no larger.

The error in estimating $\Delta M_z$, will dominate the error in
$\Omega_\Lambda$ because it is more difficult to determine precisely
and they are terms that are approximately linear in $z$.  However, for
a 20 cluster, 100 galaxies per cluster survey (plus the accompanying
field galaxies) we expect that the error in $\Omega_\Lambda$ should be
about $2/3$ of the error in $\Delta M_z$. This is shown in fact to be
borne out fairly accurately in the full nonlinear result, shown in
Figure~\ref{fig:lame}.

Differential merging has no effect on the measurement of the cluster
luminosity, $L$, or the field luminosity density, $j$, unless
accompanied by star formation.  Measurements of the [OII] line in high
luminosity cluster and field galaxies (\cite{balogh}) find that the
star formation in the field adds little mass to these galaxies, that
there is no increase in star formation upon cluster entry, and
confirms the well known suppression of star formation in clusters.  On
the other hand, merging will increase the $M_\ast$ in a way that could
be mistaken for luminosity evolution.  The colors and morphological
types of galaxies, both suitably adjusted for fading in a cluster, can
test for luminosity evolution as opposed to merging.  The current
measurements of the radial change from cluster center to pure field of
the mean luminosity and color suggest that relatively little merging
of field galaxies relative to cluster galaxies occurs. The dominant
effect is that galaxies largely cease forming stars and fade a few
tenths of a magnitude when they enter the cluster
(\cite{a2390,profile,balogh}).  The accuracy to which differential
luminosity evolution and differential merging can be determined in a
multi-color survey is mainly the precision to which the characteristic
$M_\ast$ can be measured, which we have taken as our error estimate.
The differential luminosity evolution is already known to be no more
than half of the expected variation of $\Omega_e(z)$ (if the universe
is flat). Hence, the galaxy sample will be sufficient to increase the
precision of the differential evolution measurement to allow confident
$\Omega_\Lambda$ estimation.

\section{Conclusions}

The classical volume-redshift test, which depends upon an absolute
comparison of galaxy numbers or luminosities at different redshifts,
can be modified to create a much more reliable, completely
differential, test. The extra ingredient is to combine quantities
which together give the mean co-moving mass density of the universe,
which is a conserved quantity. This benefit comes at the modest cost
of a 25\% reduction in $\Omega_\Lambda$ sensitivity.  Monte Carlo
simulations show that a sample of 2000 cluster galaxies and a
comparable field sample will be able to tightly constrain the
differential evolution between cluster and field and will measure
$\Omega_\Lambda$ to a precision of about 25\%. Moreover, groups of
galaxies found within a large field survey will serve the same purpose
and provide a second avenue to address differential evolution between
cluster and field. Differential evolution between clustered and field
galaxies will be addressed using multi-color photometry and imaging.
In the longer term, a survey of 100 or so rich clusters will increase
the precision of the geometry measurement to about 7\%. Such data can
also give extremely precise measurements of the evolution of the
sample galaxies, although these should not be taken as absolute
evolutionary measurements unless care is taken to avoid redshift
dependent selection effects.

\acknowledgments

I thank Howard Yee for insightful discussions. This research was
supported by NSERC of Canada.

\clearpage

\figcaption[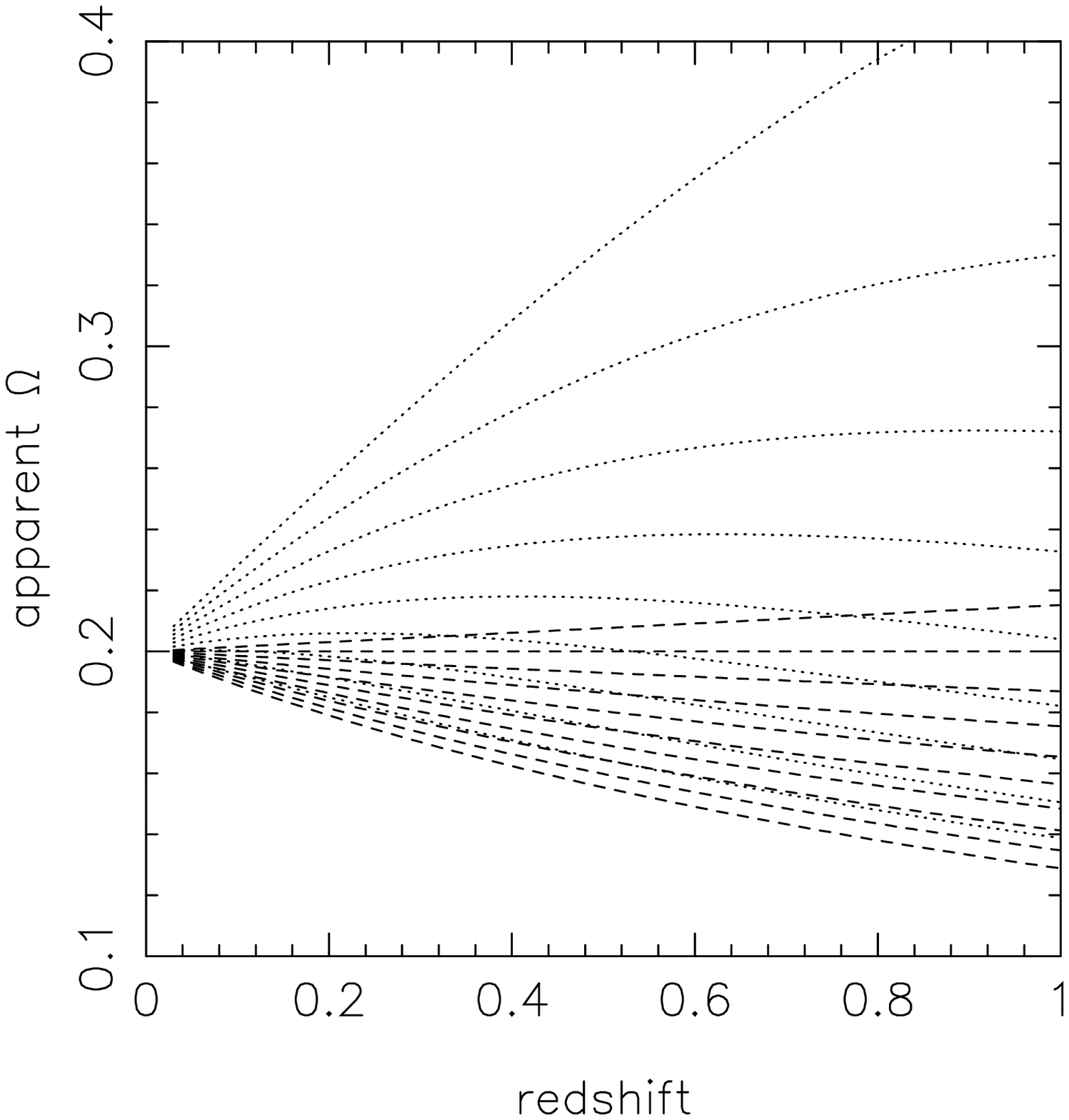]{
The variation of the apparent $\Omega_e(z)$ with redshift, where we
have assumed that $\Omega_M^i=0.2$ and $\Omega_\Lambda^i=0$. The true
values of $\Omega_M$ range from 0.1 to 0.9 (top to bottom) for
$\Omega_\Lambda=0$ models (dashed lines) and flat,
$\Omega_M+\Omega_\Lambda=1$, models (dotted lines).
\label{fig:lamm}}

\figcaption[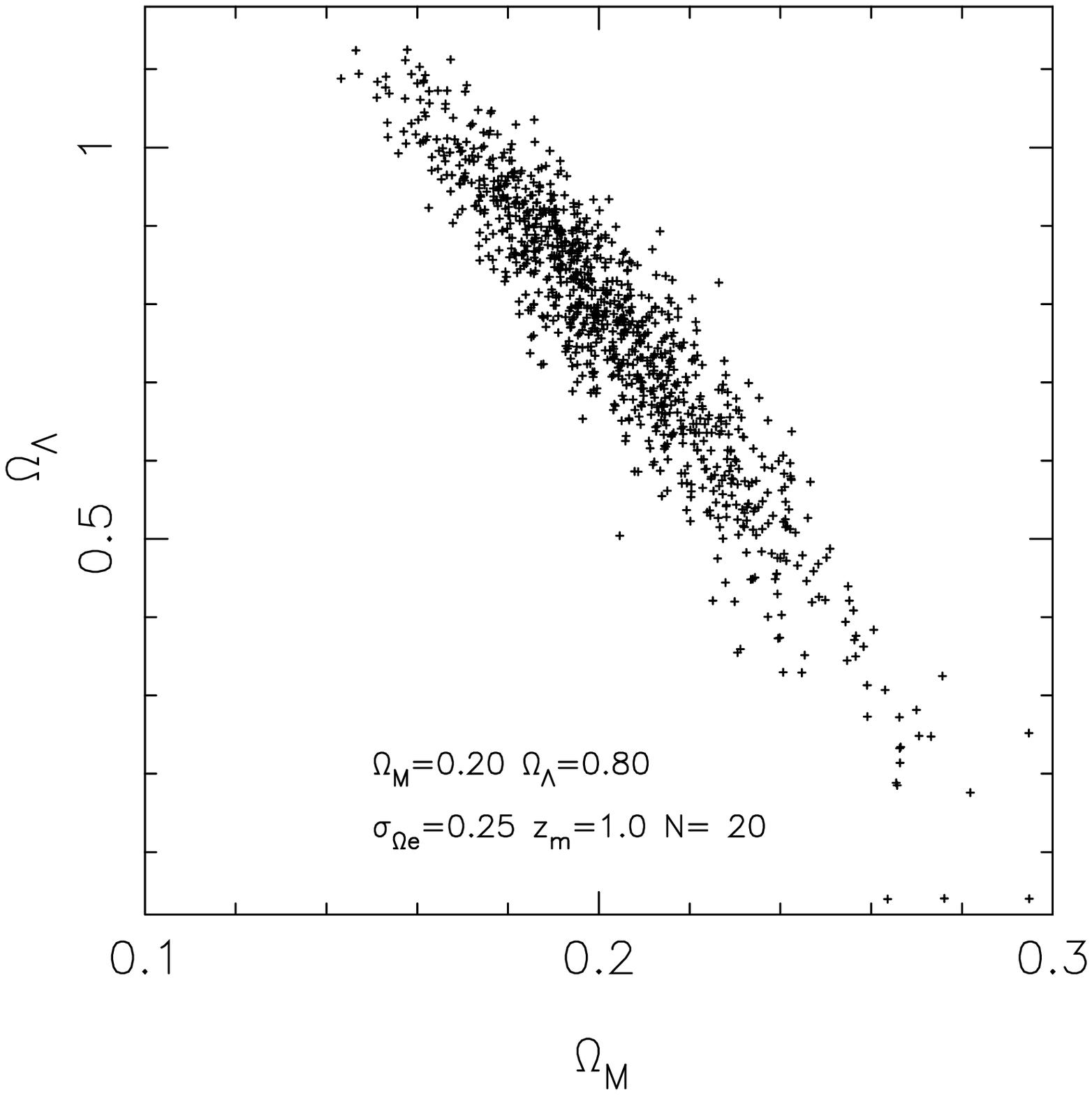]{
The 68, 90 and 99\% confidence contours in $[\Omega_M,\Omega_\Lambda]$
for a sample of 20 clusters between redshift 0 and 1 in a random
uniform distribution, with 25\% statistical errors in their
$\Omega_e(z)$ values. The true model is $[0.2,0.8]$.  This sample
would contain 2000 cluster galaxies. A greater than 99\% confidence
discrimination between flat and open low density models can be made.
\label{fig:chi20}}

\figcaption[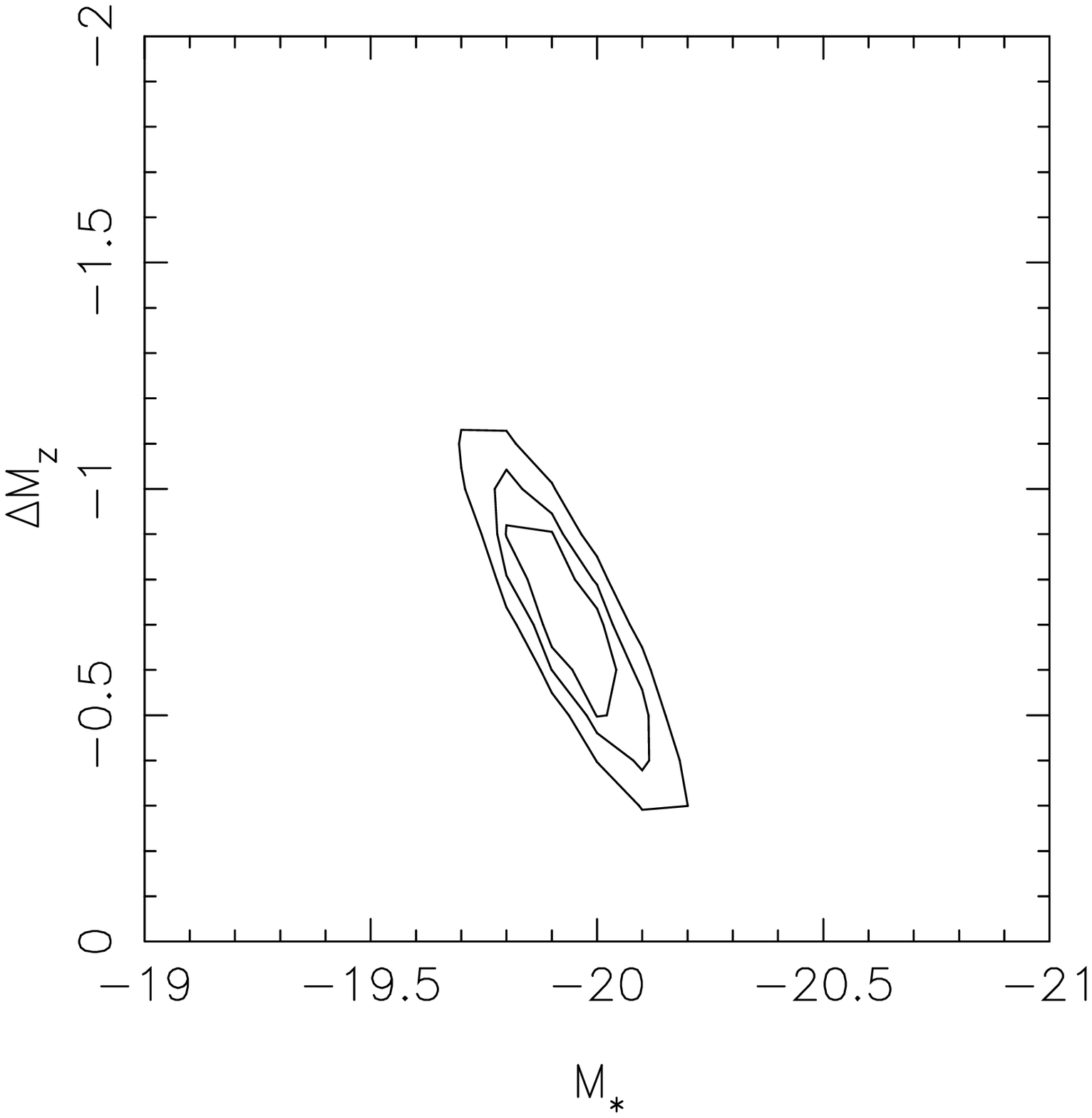]{
The 68, 90 and 99\% confidence contours in $[M_\ast,\Delta M_\ast]$
for the measurement of the luminosity function and its evolution. The
error in determining $\Delta M_\ast$ is about 0.2 mag.
\label{fig:chilf}}

\figcaption[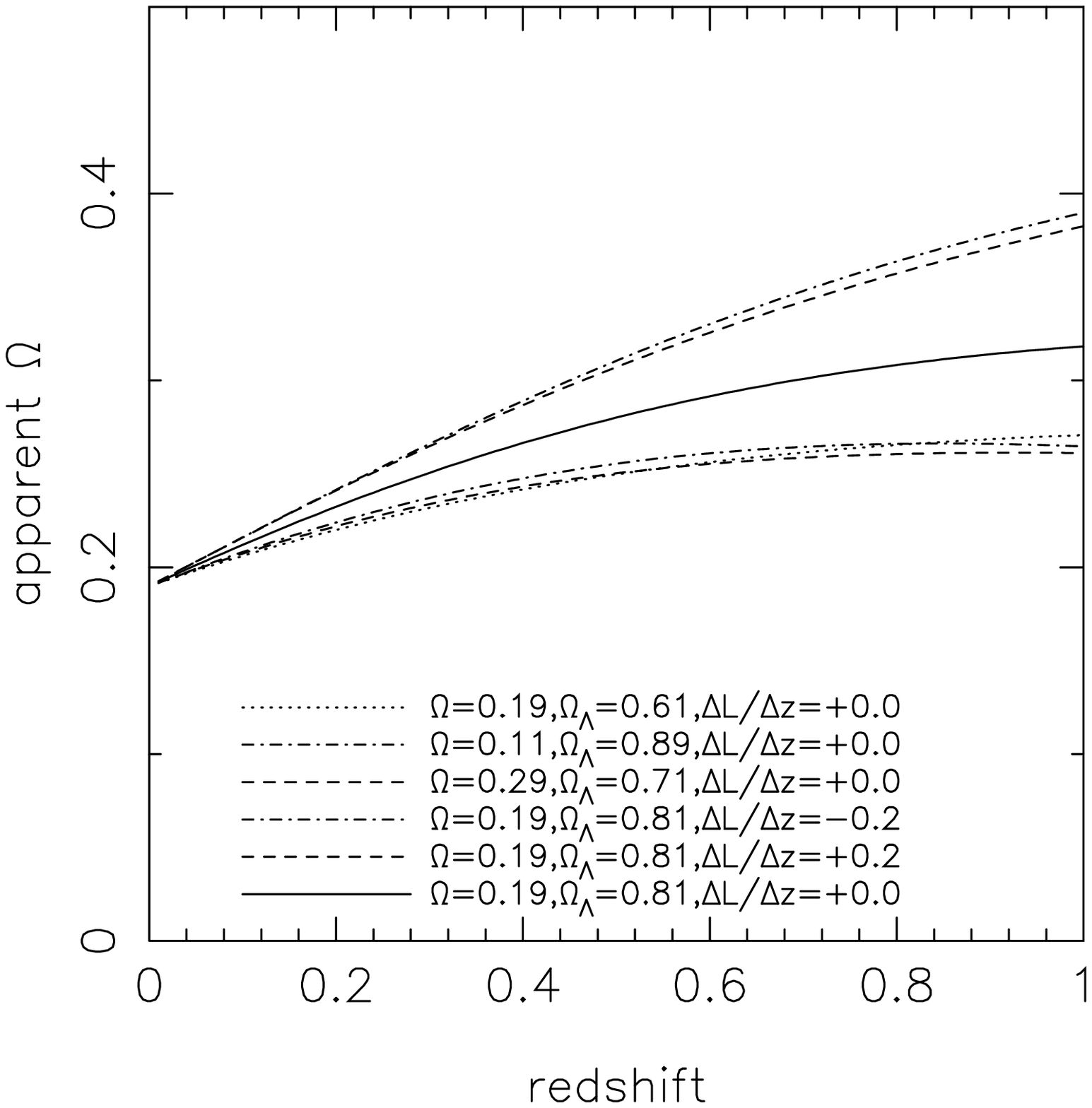]{
The effect of errors in the rate of differential evolution of cluster
and field galaxies.  The error in $\Omega_\Lambda$ is approximately
two-thirds of the error in measuring the $\Delta_M$.
\label{fig:lame}}

\begin{figure}[ht] \figurenum{1}\plotone{fig1.ps} \caption{}\end{figure}  
\begin{figure}[ht] \figurenum{2}\plotone{fig2.ps} \caption{}\end{figure}  
\begin{figure}[ht] \figurenum{3}\plotone{fig3.ps} \caption{}\end{figure}  
\begin{figure}[ht] \figurenum{4}\plotone{fig4.ps} \caption{}\end{figure}  

\end{document}